# Interacting chains of orbital polarons in "Colossal" magnetoresistive La$_{1-x}$Sr$_x$MnO$_3$ revealed by spin and lattice dynamics

M. Hennion[*] and S. Petit
*Laboratoire Léon Brillouin, CEA-CNRS, Université Paris-Saclay, CEA Saclay, 91191 Gif-sur-Yvette, France*
A. Ivanov
*Institut Laue-Langevin, 71 avenue des Martyrs, 38000 Grenoble, France*
and
*Peter the Great St.Petersburg Polytechnic University, 195251 St.Petersburg, Russia*
J. P. Castellan and D. Lamago
*Institut für Festkörperphysik, Karlsruher Institut für Technologie, D-76021 Karlsruhe, Germany*

[*] E-mail: martine.hennion@cea.fr

**Abstract**

The origin of the effect of "colossal magneto-resistance" (CMR) remains still unexplained. In this work we revisit the spin dynamics of the pseudo-cubic La$_{1-x}$Sr$_x$MnO$_3$ along the Mn-O-Mn bond direction at four *x* doping values (*x*<0.5) and various temperatures and report a new lattice dynamics study at $x_0$=0.2, representative of the optimal doping for CMR. We propose an interpretation of the spin dynamics in terms of orbital polarons. This picture is supported by the observation of a discrete magnetic energy spectrum $E^n_{mag}(q)$ characteristic of the internal excitations of "orbital polarons" defined by Mn$^{3+}$ neighbors surrounding a Mn$^{4+}$ center with a hole. Because of its hopping, the hole mixes up dynamically all the possible orbital configurations of its surrounding Mn$^{3+}$ sites with degenerate energies. The $E^n_{mag}$ values indicate a lift of orbital degeneracy by phonon excitations. The number *n* varies with the spatial dimension *D* of the polaron and the *q*-range determines its size. At *x*=0.125 and *x*=0.3 the spectrum reveals 2*D* polarons coupled by exchange and 3*D* "free" polarons respectively, with sizes *l*=1.67*a* < 2*a* in all bond directions. At $x_0$=0.2, the spin and the lattice dynamics provide evidence for chains of orbital polarons of size *l*=2*a* with a periodic distribution over ≈ 3*a* and an interaction energy ≈ 3 meV. At $T \leq T_c$ the charges propagate together with the longitudinal acoustic phonons along the chains enhancing their ferromagnetic character. The phase separation between metallic and ferromagnetic chains in a non-metallic matrix may be crucial for CMR.

## I Introduction

In the rich variety of strongly correlated quantum materials[1-5] manganites are known for their "colossal" magneto-resistance properties (CMR). The importance of nanoscale magnetic inhomogeneities observed close to the transition temperature T$_c$ to the ferromagnetic and metallic phase was early recognized[6-9]. The role of the "correlated lattice polarons" observed at T>T$_c$ was highly debated[10-13] and phonon anomalies were reported in the high energy range[14-16]. However, the true origin of the CMR property remains obscure. In this context, the study of spin dynamics has attracted a lot of attention. Below T$_c$, in all the compounds, the magnetic excitation spectrum is unconventional with a significant broadening and softening of the spectra close to the zone boundary (*q*=π/*a* where *a* is a pseudo cubic lattice parameter)[17-22]. Their magnetic origin was demonstrated by polarized neutrons[23]. A classification of the compounds was proposed based on the width of the electronic band, depending on the existence of charge-ordered structures in the phase diagram[24]. In this paper, we focus on the La$_{1-x}$Sr$_x$MnO$_3$ compound with the largest electronic band. The maximum of the magneto-resistance occurs at $x_0$=0.17 [25].

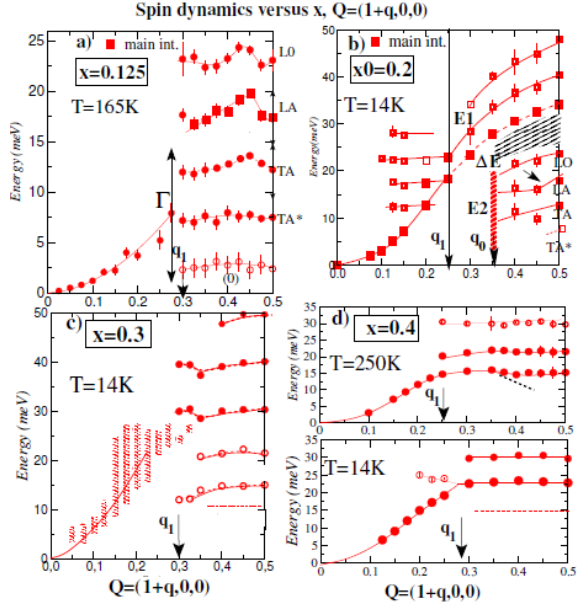

**Fig 1.** Spin dynamics spectra of $La_{1-x}Sr_xMnO_3$ along Mn-O-Mn bond direction. The $q_1$ value indicates a limit of the discrete energy spectrum. At $x_0=0.2$ two q-limits appear, defined by $q_0 \approx 0.35$ rlu and $q_1 \approx 0.25$ rlu. At $x=0.125$ and $x_0=0.2$, the square symbols correspond to the excitations with main intensity. The continuous lines are guides for the eyes. The pattern styles are unresolved excitations. **a)** $x=1/8$, T=165K. The empty circles of the lowest energy level (o) are temperature dependent (see text). $\Gamma$ shows the full energy width at half maximum **b)** $x_0=0.2$, T=14K. The hatched area indicates an energy separation between the E1 and E2 ranges. **c)** $x=0.3$, T=14K. **d)** $x=0.4$, T=14K (lower) and T=250K (upper). The black dotted line shows the tendency for Brillouin zone folding.

Previous inelastic neutron scattering have shown that in the large q-range, close to the zone boundary, the spin wave broadening can be resolved into a series of n discrete modes. They are called in this paper $E^n_{mag}(q)$[26,27]. Importantly, in the charge-ordered state observed at $x=0.125$, T< $T_{o'o''}$ ($T_{o'o''}$=159K is the charge-ordering transition), the dispersion of the magnetic branches was found in coincidence with the phonon ones[28]. In this work, we revisit the spin dynamics of $La_{1-x}Sr_xMnO_3$ for several x doping values, ranging from the quasi-metallic to the metallic states[29]. In addition to the three compounds previously studied[26,27], namely with $x=0.125$ for T> $T_{o'o''}$, $x_0=0.2$, representative of the optimum doping for the CMR, and $x=0.3$, we report a new determination of the spin excitation spectrum at $x=0.4$. On this basis, we propose a comprehensive analysis of the discrete spectrum as a function of doping. We also determine the lattice excitations spectrum in the acoustic range for $x_0=0.2$, around $T_c$, providing a complete picture of the magnetic and lattice excitations at the optimal magneto-resistance doping value.

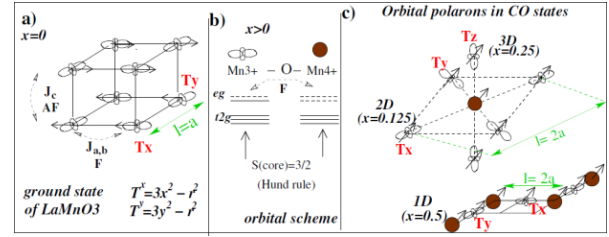

**Fig 2.** a) Staggered pattern of the orbital states $T^x$ and $T^y$ in the (a, b) planes of the pseudo-cubic $LaMnO_3$. It determines an F spin coupling in (a,b) planes and an AF one along c according to the Goodenough-Kanamori rules. **b)** Scheme of the orbital states of the $Mn^{3+}$ and the $Mn^{4+}$ ions. **c)** Evolution with x of the shape or the dimension D of the orbital polarons taken from Mizokawa et al (Reference 24) in the charge ordered (CO) states of manganites.

Our present analysis is based on the two following observations. First, as shown in Fig 1, the $E^n_{mag}(q)$ spectrum appears in a well-defined q-range above the specific value $q_1$ indicated by the arrows in Fig.1. Below $q_1$, the spin excitation spectrum is essentially represented by the quadratic spin wave dispersion law $E \sim Dq^2$, with the stiffness constant **D**, typical of the ferromagnetic metallic state at T<$T_c$. This defines a wave vector scale which separates "collective" or "global" magnetic excitations for $q < q_1$ from "individual" or "local" ones for $q > q_1$. In the intermediate $q_1/2 < q < q_1$ range, the two types of excitations coexist. This is especially well observed for $x_0=0.2$ and $x=0.3$.

Next, we follow the idea that the discrete magnetic spectrum $E^n_{mag}(q)$ results from the existence of "orbital polarons" in a

(quasi)metallic state. Such a polaron has been introduced in manganites because of the anisotropic shape of the $d$ orbital. It consists of a cluster of neighboring $Mn^{3+}$ sites whose $d$ orbitals point towards a central $Mn^{4+}$ ion occupied by the hole[24]. In charge-ordered states, the polaron is ferromagnetic (F), Fig 2-c. To describe this picture, we remind that each $Mn^{3+}$ ion hosts four electrons (see the scheme in Fig 2-b). Three of them are in a $t^{2g}$ orbital state and, because of the Hund's rule, form an electrically inert core spin $S=3/2$. The outer one is in a Jahn-Teller active orbital state corresponding to either $x^2-y^2$ ($T^z$) or $3x^2-r^2$ ($T^x$), $3y^2-r^2$ ($T^y$) and $3z^2-r^2$ ($T^z$) states, with a spin $s=1/2$ also aligned with the core spin. In the parent orthorhombic $LaMnO_3$, a cooperative Jahn-Teller effect lifts the orbital degeneracy by forming a staggered pattern of $T^x$ and $T^y$ components, from which the magnetic structure of ferromagnetic $(a,b)$ planes stacked antiferromagnetically along $c$ is obtained via the Goodenough-Kanamori rules (Fig 2-a). The $E^n_{mag}(q)$ spectrum appears upon doping, as the antiferromagnetic (AF) exchange $J_c(x)$ vanishes and the quasi-metallic state arises. There, as the hole at the center $Mn^{4+}$ of the polarons is hopping, the orbitals of the $Mn^{3+}$ neighbors fluctuate to accommodate with the new position of the hole. Equivalently, the hole mixes up dynamically all the possible orbital configurations of its $Mn^{3+}$ neighbors with degenerate energies[30]. In these compounds where the orbital-phonon coupling is expected to be strong[31], the phonons lift the orbital degeneracy through a dynamical Jahn-Teller effect. Some constraints on the hole hopping[32] and on the orbital structure of the polarons suggest that the $T^x$, $T^y$, $T^z$ orbital states are mainly concerned with this mixing. In the case where all sites are $Mn^{3+}$ ions, the interplay between the spin and orbital operators arises from the crossed term $J_s J_t \Sigma_{ij} (S_i S_j)(T_i T_j)$ of the Kugel-Khomskii Hamiltonian which in a mean field approach is averaged to $<S_iS_j>T_iT_j$[33,34]. It corresponds to orbital excitations with an intensity proportional to the magnetization ($<S_iS_j> \infty M(T)$). In metallic state, one of the two (i, j) site is a $Mn^{4+}$ ions, with a permanent change of configuration. Experimentally, two situations are observed. In the metallic state (x=0.3 and x=0.4 at T=14K) the excitations are independent of temperature and wave vector and their relative intensity follows the magnetization M(T). In this limit, there is no direct exchange coupling between spins or no correlations between orbital polarons, and the magnetization arises from the collective spin coupling effect. In contrast, at lower doping (x=0.125 and x0=0.2) and at larger doping close to x=0.5 (our data taken with x=0.4 at T=250K), the energy levels exhibit dispersion. A spin coupling is superimposed over the effect of orbital fluctuations which reveals the existence of correlations between the orbital polarons.

The splitting of the orbital degeneracy by phonons allows one to determine the number of orbital degenerate states $n$. It corresponds to the number of possible orientations of the orbital states involved in the orbital polarons. It varies therefore with the dimension $D$ of the orbital polarons. Its evolution with $x$ has been shown in the case of the charge-ordered states[24] as illustrated in Fig 2-c. At small $x$ values ($x<0.2$), the orbital polaron is essentially a "$2D$" object with four $Mn^{3+}$ sites surrounding a central $Mn^{4+}$ ion, as a consequence of the 2D orbital ordering observed at x=0. At larger doping (x=0.25), it becomes "$3D$" with six $Mn^{3+}$ sites surrounding a central $Mn^{4+}$ ion. It shrinks beyond this doping value because of steric hindrance so that, at x=0.5, specific zig-zag paths of orbitals set in with two $Mn^{3+}$ sites surrounding a central $Mn^{4+}$ site. In this magnetic structure the zig-zag chains are ferromagnetic with an AF inter-chain coupling[35]. Experimentally, at $x$=0.125 (2$D$), $n$=4. As argued below the lowest energy level being temperature dependent should not be numbered. It may correspond to the ($x,x$), ($x,y$), ($y,x$), ($y,y$) fluctuations of orbital orientations. The number increases to $n \approx 8$ at $x$=0.3 in the 3$D$ metallic state where 9 possibilities are expected for the fluctuations between the $x$, $y$, $z$ orbital orientations, and decreases to $n$ =2 $\approx$ 3 at $x$=0.4 in agreement with the expected evolution.

In this direct space picture, the characteristic $q$-scale of the excitations defined by $q_1$ can be transformed back to real space to get the size of the polaron via the relation $l$=0.5 $a/q_1$. Here, the minimum $q$-scale value ($q_1$=0.5) is related to the lattice spacing $a$. This relation is valid as long as the excitations are localized (or $q$-independent) and originate from a coupling with the lattice (here with the phonons). This relation tells us that for $q_1 \approx$ 0.25 rlu, the size of the cluster along Mn-O-Mn directions is close to 2 lattice spacings.

Within this analysis, the spin dynamics reveals the existence of orbital polarons which evolve with $x$ into three steps $x < x_0$, $x \approx x_0$ and $x > x_0$. For $x \neq x_0$ only one $q$-scale is observed which defines the size of the polaron close to 2$a$ either coupled ($x$=0.125) or uncoupled ($x$=0.3) whereas at $x \approx x_0$, two $q$-scales are observed, both in the spin and the lattice dynamics characteristics of a local "ground state" defined by a commensurate size of polarons (2$a$) and their periodic distribution ($\lambda \approx$ 3$a$).

The paper is organized as follows. Section **II-1** describes the spin dynamic spectrum within the new analysis at $x$=0.125 and $T_{o'o''}$ <$T$<$T_c$. Section **II-2,** devoted to $x_0$=0.2, presents both the spin dynamics spectrum within the new analysis for $T \leq T_c$ in **(a)**, and the lattice excitations in the acoustic range for $T$>$T_c$ and $T$<$T_c$ in **(b)**. Section **II-3** describes the spin spectrum within the new analysis in the metallic state, at $x$=0.3 and $x$=0.4 ($T$<$T_c$). Section **III** provides the discussion and conclusion.

Inelastic neutron scattering experiments were carried out at the Laboratoire Leon Brillouin using cold and thermal neutron three axis spectrometers (TAS) 4F1, 1T, 2T and at the Institut Laue Langevin using thermal neutron TAS IN8. The instrument configuration and resolution was in each case adapted to the studied momentum and energy range. The crystal structure of the studied phases is either weakly orthorhombic ($x$=0.125, $x_0$=0.2) or rhombohedral ($x$=0.3, 0.4) averaged for simplicity to a pseudocubic structure with $a \approx$ 3.9Å. Only the Mn-O-Mn bond direction or [100] symmetry direction is considered here. The wave-vector $q$ is expressed in reduced lattice units r.l.u., defined by $q$=$q$ (2$\pi/a$,0,0) along the symmetry direction [100]. At $x_0$=0.2, the phonons are distinguished from magnons by changing the Brillouin zone $\tau$ in the total wave vector $Q$ where $Q$=$q$+$\tau$ and by using the effect of temperature ($T$>$T_c$).

## II Experiments

### II-1 The quasi-metallic regime, $x$=0.125 ($T$>$T_{o'o''}$ =159K, $T_c$=181K)

The spin dynamics spectrum reported in Fig 1-a has been measured in the whole $q$-range at $T$=165K ($T_{o'o''}$ <$T$<$T_c$), and at one $q$-value ($q$=0.5 rlu) as a function of temperature up to 250K along Mn-O-Mn and up to 292K in other symmetry directions. In small-$q$ range**,** the stiffness constant $D$ (global coupling) has a 2$D$ character[26]. The acoustic phonon excitations have been also measured but not reported here. An example of raw data is shown at $q$=0.15 rlu in Fig 3-a. The fast increasing damping $\Gamma$ of the excitations which follows a $\Gamma$=A $q^{2.5}$ law at T=165K smears the transition between the small-$q$ and the large-$q$ regimes (Fig 3-c). As shown from the raw data previously reported for four $q$ values[26], the $E^n_{mag}(q)$ spectrum consists of 4 equidistant levels separated by $\approx 5$ meV, namely 22.5, 17, 12, 7.5 meV at $q=0.5$ rlu. The lowest energy level, $E(0) \approx 2.5$ meV, is not considered. It is actually temperature dependent since it disappears in the "charge-ordered" state ($T < T_{o'o''}$) in contrast with the four $q$-dispersed energy levels which are in coincidence with the $q$-dispersed phonons[28]. It looks like therefore as an anisotropy gap of magnetic and possibly structural origin and should not be considered when numbering the energy levels related to orbital degeneracy.

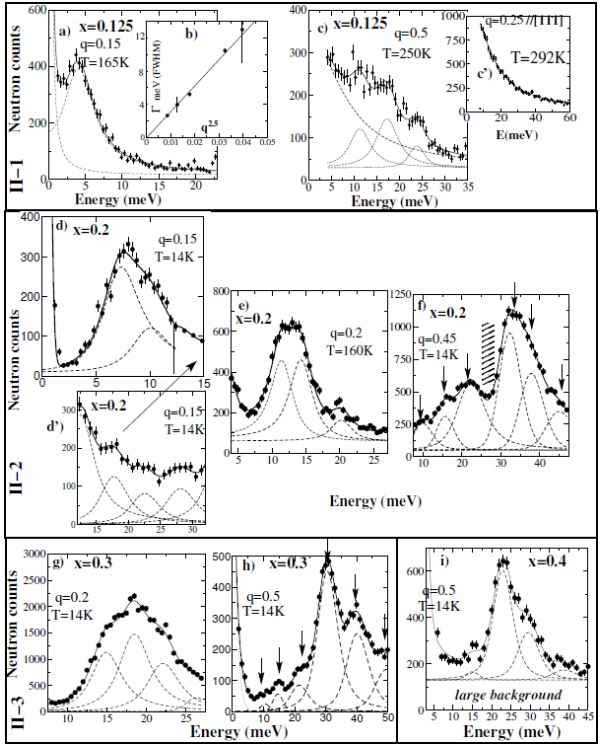

**Fig 3**. Raw data corresponding to the spin dynamic spectra $E(q)$ in the three regimes **II-1, II-2, II-3** at small $q < q_1$ and large $q > q_1$ $q$ values. The wavevector $q$ is expressed in relative lattice units (rlu). The data are fitted by Lorentzian line-shapes to account for damping, convoluted with the resolution function of three-axis spectrometers. In the large-$q$ range, the intrinsic energy linewidth of each energy level (full width at half maximum) is $\Gamma \approx 4{-}6$ meV. **II-1:** $x=0.125$. **a)** $q=0.15$ T=165K ($T_{o'o''} < T < T_c$). **b)** $q=0.5$, T=250K ($T > T_c$) Inset **b'**): $q=0.25$, T=292K along [111]. **c)** $q^{2.5}$ dependence of the energy linewidth $\Gamma$ of the small $q$-range excitations. **II-2:** $x_0=0.2$. **d)** $q=0.15$, T=15K with the tail of energy modulations in **d'**, **e)** $q=0.2$, T=160K for a direct comparison with $x=0.125$, **f)** $q=0.45$ T=15K. The pattern style corresponds to $\Delta E$ (see the text). **II-3: g)** $x=0.3$ $q=0.2$, T=14K, **h)** $x=0.3$ $q=0.5$, T=14K. The temperature dependent spectra have been reported in Reference 27. **i)** $x=0.4$, T=15K.

The energy $E^n_{mag}(q)$ values observed at $q=0.5$ rlu are equal or very close to the $q=0.5$ rlu phonon excitations LO, LA and TA reported in Fig 4 for $x_0=0.2$. The value $E \approx 7.5$ meV is related to a TA* branch, specific of $x=0.125$ value. From the $q_1=0.3$, a "size" of polarons $l \approx 1.7a$ ($l=0.5\, a/q_1$) is deduced along the Mn-O-Mn bond directions, indicating a strong lattice contraction. The most surprising observation here is the existence of a $q$-dispersion with a maximum of energy at $q=0.45$ rlu. This is particularly well observed for the energy level $E \approx 17$ meV with strongest intensity. Such a $q$-dependence may be interpreted as an AF coupling between two neighbor F polarons, each of them being on a contracted lattice scale $l \approx 1.7a$. It correspond to two polarons (or bipolarons) coupled in a singlet state. Actually the incommensurability provides a complexity to the analysis which leads us to explain this observation in a further work. The discrete spectrum is still observed at 250K, well-above $T_c$ (181K) superimposed on a quasi-elastic signal, typical of a paramagnetic state. Corresponding raw data are reported in Fig 3-b. It is no longer observed at 292K. This is illustrated in the inset **b'** of the Fig 3-b along the symmetry direction [111] in agreement with NMR experiments[35].

**II-2 The transitory regime: $x_0=0.2$, $T_c=301$K with optimal magnetoresistance**

**a) The spin dynamics spectrum**

In Fig 1-b at T=14K the spin dynamics spectrum exhibits a change in energy scale by a factor 2 with respect to that observed at $x=0.125$. Raw data are reported in Fig 3-d-d'-e-f. In the large-$q$ range, at $q=0.35$ rlu, an interval $\approx 8$ meV appears which is larger by $\Delta E \approx 3$ meV than the interval $\approx 5$ meV of the lower energy range. It is indicated by the hatched area in Fig 1-b and in Fig 3-f. This $\Delta E$ ($q=0.35$ rlu)=3 meV separates two zones of discrete spectrum $E2$ ($\leq 25$ meV) and $E1$ ($\geq 25$ meV) with two typical $q$-dependences. $E2$ with 4 energy levels in the same energy range as $x=0.125$, corresponds to the "in-plane" or $(x,y)$ orbital fluctuations, so that $E1$, at higher energy is attributed to new "out-of-plane" ones. The energy $\Delta E$ between the two zones $E1$ and $E2$ indicates the existence of a connection between the "in-plane" and the "out-of-plane" orbital fluctuations. At T=14K, the zone $E2(q)$ appears at $q_0 \geq 0.35$ rlu where the discrete levels follows the $q$-dependent phonon branches. This can be in particular verified for LA($q$) where the anomaly in the $q$-dispersion indicated by an arrow at $q=0.45$ rlu in Fig 1-b is also observed in the LA*($q$) branch reported below in Fig 5-a. Such a coincidence reminds the $q$-dependent magnon-phonon coincidence previously observed at $x=1/8$, T< $T_{o'o}$"[28]. It indicates the existence of strong orbital correlations along the Mn-O-Mn bond directions in the plane, consequence of a charge ordering with a cut-off value $q_0 \approx 0.35$ rlu. This value determines a periodic distribution of charges with a period $\lambda_0 = 2\pi/q_0 \approx 3a$.

The higher energy $E1(q)$ zone may be analyzed as the superimposition of two spectra of distinct origins. One, $q$-independent, spreading on the [0.25-0.5] $q$-range, determines links of $2a$ size along the $z$ direction perpendicular to the $(x, y)$ planes. The other one, $q$-dependent, defines a cosine law which approximately fits to the $Dq^2$ law defined in the small-$q$ range. This $q$-dependence is observed actually in the whole temperature range $T \leq T_c$. This is shown in Fig 4-a where a T-dependent variation can be drawn through the energy values, corresponding to $D$(T) (stiffness constant) or to the magnetization M(T). Therefore, the magnetic excitations of the large-$q$ range related to the orbital correlations along $z$ exhibit a collective character which arises from the same ferromagnetic state as the magnetic excitations of the small $q$-range. Finally, a quantized spectrum with a weak intensity is also observed in the $q$-range of the "collective" excitations above the quadratic $Dq^2$ down to $q=0.125$ rlu (Fig 1-b) with raw data in Fig 3-d' and Fig 3-e. It defines a $4a$ scale as for two orbital polarons of $2a$ size in contact one to the other. No further scale can be observed which is consistent with the situation in metals where as $q$ goes to zero, a sum rule of intensity exists in favor of the Goldstone mode ($q=0$, $E=0$) of the collective excitations.

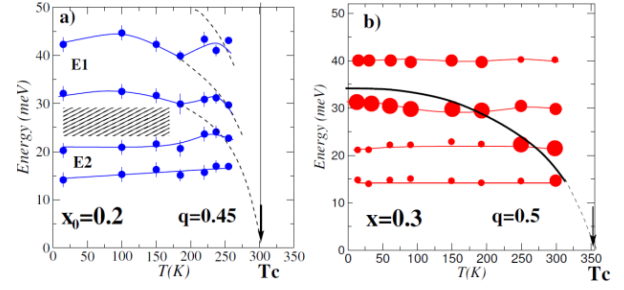

**Fig 4.** The temperature evolution of the four central energies of the $E^n_{mag}$ spectrum **a)** at $x_0=0.2$ for $q=0.45$ rlu and in **b)** at $x=0.3$ for $q=0.5$ rlu. The dashed and continuous lines are guides for the eyes. In **a)** the black pattern indicates the separation in energy scale: between the low-energy excitations, temperature-independent, with energy values equal to the phonon energies measured at q=0.5 rlu reported in Fig 5, and the high-energy ones, temperature-dependent, which follow variation of $D$(T) or the magnetization M(T). In **b)**, the distribution of the intensity over the spin spectrum dynamics is shown by circles of varying size. The black line indicates the variation of the energy with the strongest intensity as a function of temperature while approaching $T_c$.

From these observations, a picture of phase separation into ferromagnetic chains arises. The chains consist of "orbital polarons"

of 2*a* size with a periodic distribution ≈ 3*a* in the perpendicular bond direction of the plane which contains the chains. This is a 2*D* picture. To fill the 3*D* space, we have to assume the existence at time "*t*" of several equivalent planes with a weak interaction, distant each from the other from a single lattice parameter. The two bond directions perpendicular to the chains become equivalent only in space and time average. Such a picture leads to the doping value $x=1/6 \approx 0.17$ for which the CMR property is optimum in this compound[25]. This picture is supported by the interaction with the lattice described below.

**b- The lattice dynamics spectrum**

Fig 5-a displays low-energy ($E \leq 25$ meV) phonon excitations with longitudinal and transverse character obtained at T=350K (T>$T_c$) and T=290K (T<$T_c$) along Mn-O-Mn. Corresponding raw data are displayed in Fig 5-c-d. The transverse and the longitudinal excitations have been obtained by using distinct experimental configurations. Here, the scattering of magnetic origin is a background in the raw data with intensity reduced by the form factor of Mn ions. At T=350K, a nearly constant-energy value, $E \approx 8$ meV, is observed in the [0.25-0.5] *q*-range, labeled TA*. It is connected with the TA(*q*) branch of the 3*D* pseudo-cubic structure at $q_1=0.25$ rlu. At lower energies, a dispersed TA$^{perp}$(*q*) branch is observed up to $q_0=0.35$ rlu, where a small shift in energy appears in TA*, defining $\Delta E$ ($q_0=0.35$ rlu)=3 meV. At T=290K < $T_c$, the TA$^{perp}$(*q*) branch has nearly disappeared. In contrast, the TA* one persists with the same behavior at $q_0=0.35$ rlu and with an increasing intensity at the *q*=0.25 rlu value. Such increase is in contradiction with the Bose factor. It reminds previous observations reported at higher energy[15,16]. Concomitantly new

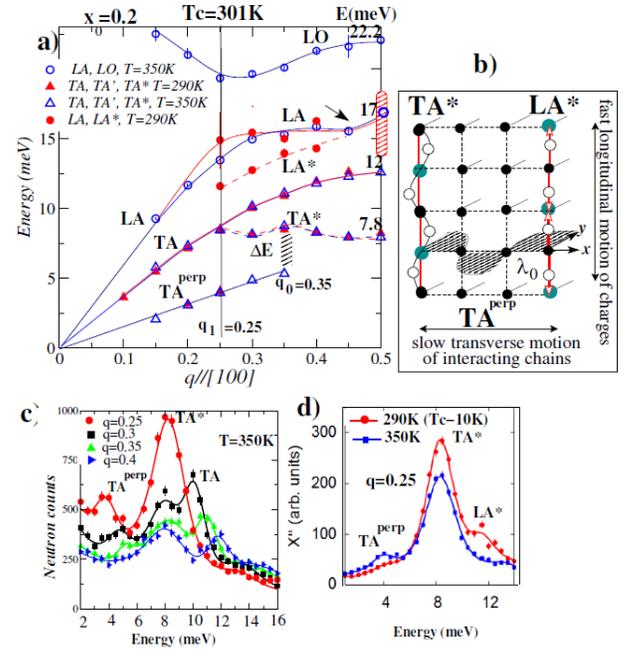

**Fig 5 a)** Lattice dynamics spectrum ($E \leq 25$ meV) for T=350K (full symbols) and T=290K (empty symbols) measured at $x_0=0.2$. The red pattern style at *q*=0.5, $E \sim 15$ meV indicates a broadening induced by the interaction between the LA and LA* excitations. The arrow in the LA branch indicates an anomaly in energy also observed in the quantized spin spectrum of **Fig 1-b**. The black pattern style defines the interaction energy $\Delta E=3$ meV at $q_0=0.35$ rlu between the TA* and TA$^{perp}$ branches.

**b)** Instantaneous picture of interacting chains of ordered "orbital polarons" of size 2*a* in one of the *(x,z)* planes. Chains are stretched along *z* direction and separated by ≈3*a* in the *x* direction. Planes are supposed to be not correlated in *y* direction. The blue circles correspond to Mn$^{4+}$, the black ones to Mn$^{3+}$ and the empty ones to O atoms. The TA* and LA* branches correspond to acoustic excitations with respectively a transverse and stationary character or a longitudinal and propagating one with wave vector along the chains. The TA$^{perp}$ branch corresponds to slower and propagating transverse fluctuations of at least two interacting chains with a distribution defined by $q_0 < 0.35$ rlu or $\lambda > \lambda_0 = 2\pi/q_0 \approx 3a$.

**c)** Raw data corresponding to the transverse acoustic excitations as a function of *q* for T>$T_c$, showing the disappearance of TA$^{perp}$ at *q*=0.4 rlu. **d)** Raw data corresponding to *q*=0.25 at T=350K > $T_c$ and T=290K < $T_c$ corrected by the Bose factor.

excitations occur in the [0.25-0.5] *q* range, such as the dispersed LA* branch just below the usual LA(*q*) branch which consequently is shifted to an higher energy values at *q*=0.25 rlu. These observations are analyzed as follows.

The existence of two transverse acoustic branches, TA* and TA$^{perp}$(*q*), reveals a privileged direction among the three Mn-O-Mn ones, induced by the charge-phonon coupling, so that the transverse acoustic excitations with wavevectors *q* parallel to that direction (TA*) are distinct from those with wavevectors *q* perpendicular to it, (TA$^{perp}$(*q*)). An instantaneous picture of two chains with transverse excitations, TA* and TA$^{perp}$(*q*), is shown in Fig 5-b. Moreover, the TA* and TA$^{perp}$(*q*) branches are defined in two separated energy ranges so that they are characteristic of two distinct time scales of lattice vibrations. The propagating TA$^{perp}$(*q*) branch, at low energy, corresponds to slow transverse fluctuations of interacting chains within a plane. During this slow fluctuation time, faster excitations occur with a *q* parallel to the direction of the chains, either transverse and stationary corresponding to the TA* branch or longitudinal and propagating corresponding to the LA* branch. Both types of excitations are observed in the [0.25-0.5] *q*-range, which is characteristic of the 2*a* scale expected for orbital polarons in contact each with the other. The lock-in of the TA* branch at *q*=0.25 rlu, ensures the structural and electrical stability of this charge-orbital-lattice ordered structure whatever the directions of the interacting chains in 3*D* space. These features are observed even at T=350K above T$_c$ (301K). The interaction energy Δ*E*=3 meV between TA$^{perp}$ and TA* occurs at *q*=0.35 rlu. This *q*-value determines the same periodic distribution of the chains in the planes as that observed in the magnetic discrete spectrum described above. The difference however is that this periodic distribution is defined here in the complementary *q*-range *(q$_x$, q$_y$ ≤ 0.35 rlu)*, by a maximal instead of a minimal wavevector limit *q*$_0$. We can conclude that this local charge ordered structure is stabilized by a fine equilibrium between all the degrees of freedom. The occurrence of an increasing intensity of TA* at the *q*=0.25 rlu value for T<T$_c$, indicates a phase coherence between the localized TA* acoustic excitations. It appears at T≤T$_c$ concomitantly with the propagating LA*(*q*) excitation and is therefore induced by the cooperative motion of the charge carriers along the chains. Coming back to the spin dynamic spectrum, we see here the evidence that the enhancement of the ferromagnetism is induced by the cooperative mobility of the charge carriers along the chains. The TA$^{perp}$(*q*) branches is observable as long as its own characteristic time, 1/*f*, deduced from its frequency values *f*, is faster than the lifetime of the chains which actually decreases as the temperature is lowered. As shown at T=290K in Fig 5-d, the acoustic transverse fluctuations TA$^{perp}$(*q*) progressively disappear below T$_c$.

### II-3 The metallic regime *x*=0.3 (T$_c$=358K), *x*=0.4 (T$_c$=315K)

At *x*=0.3 (T$_c$=358K), the spectrum reported in Fig 1-c at T=14K with raw data in Fig 3-h extends on the [≈0.3-0.5] *q*-range. It indicates a lattice contraction (*l* ≈ 1.7*a* < 2*a*) for the "size" of the orbital polarons in all bond directions, similar to that observed in the quasi-metallic state (*x*=0.125). The energy levels are nearly *q*-independent in the [0.3-0.5] *q* scale but also temperature-independent, as seen in Fig 4-b. The corresponding raw data have been reported in Reference 27. The energy values *E*=15 meV and *E*=21 meV of the $E^n_{mag}$ spectrum (*E* < 25 meV), slightly smaller than those observed at *x*=0.125, agree with the *q*-average of the phonon energies on the *q* range of the polarons. This tight relation with phonon states reveals an absence of direct spin coupling. The magnetic character of the excitations appears in the distribution of intensity among the discrete $E^n_{mag}$ values. Although the energy values are temperature independent, their relative intensity is temperature dependent. By using the raw spectra observed at $q$=0.5 rlu reported in the figure 4 of Reference 27, the energy with maximal intensity can be reported at each temperature. It provides the black line of the Fig 4-b.

At the nominal $x$=0.4 doping value, the observed $E^n_{mag}$ spectrum displayed in Fig 1-d with raw data in Fig 3-i exhibits the same energy values as those observed at $x$=0.3 with a decrease of the $n$ value ($\approx$2 or 3 at T=14K). This evolution occurs with a decrease of the stiffness constant $D$ at T=14K and of the $T_c$ values with respect to $x$=0.3. In the upper Fig-1-d the discrete spectrum determined at T=250K spreads on the [0.25-0.5] q-scale and exhibits a tendency for folding the Brillouin zone outlined by a dotted black line. This is the consequence of the slowing down of the orbital fluctuations by approaching $T_c$ expected for a doping value close to $x$=0.5. Due to the steric hindrance, the orbital fluctuations should occur along 1$D$ paths defined by the hopping holes with $T^x$, $T^y$, and $T^z$ orbital states, close to the zig-zag paths of orbitals proposed in the charge-ordered state $x$=0.5 [24] illustrated in Fig 2-d.

### III Discussion and conclusion

The main results of our analysis are summarized as follows. At low and large doping range where $q_1$=0.3 rlu the spin spectrum reveals orbital polarons with the sizes $l \approx 1.7\ a$ along the Mn-O-Mn bonds which is characteristic of a strong lattice contraction. They appear either with a 2$D$ character and an AF inter-polaron spin coupling ($x$=0.125 in quasi-metallic state) or with a 3$D$ character and "free" from any direct spin coupling ($x$=0.3 and $x$=0.4 in the metallic state T=14K). At the intermediate $x_0$=0.2 value, the spin spectrum is richer. It reveals two $q$-ranges, [0.35-0.5] and [0.25-0.5] associated with two distinct energy ranges corresponding to "in-plane" and "out-of-plane" correlated orbital fluctuations, separated by the energy $\Delta E$=3 meV. The same $q$-ranges are observed in the transverse acoustic excitations of the lattice with a $q$ wavevector respectively along one bond and perpendicular to it. Both types of excitations reveal the existence of chains consisting of orbital polarons of size 2$a$, with a periodic distribution 3$a$ in the plane of the chains. At T $\leq$ $T_c$, the charges propagate together with the longitudinal acoustic phonons along the chains, enhancing their ferromagnetic character. This picture corresponds to a phase separation. The magnetic signature of this phase separation appears in the $q$-dispersion of the spin dynamics when considering the excitations with principal intensity (square symbols in Fig 1-b). There, the $q$-dispersion is very close to a cosine law and therefore can be fitted by using one phenomenological constant $J_1$ between the first neigbor spins. In that analysis, the small-$q$ and large-$q$ ranges of the magnetic excitations are the collective excitations of the same ferromagnetic state at $x \approx x_0$. This contrasts with the $x \neq x_0$ case where the large-$q$ excitations give rise to a hardening ($x < x_0$) or a softening ($x > x_0$) (Fig 1) so that additional phenomenological constants are required to fit the whole excitations. Moreover, in these ferromagnetic chains, the magnetic and the conduction properties are coupled together. This appears in the fact that the $q_0$=0.35 rlu wavevector value which determines the periodic distribution of the chain $\lambda \approx 3a$ appears as a minimum or a maximum wavevector value of the excitations depending on their magnetic or lattice origin. We conclude that these observations characterize a phase separation between metallic and ferromagnetic chains embedded in a non-metallic matrix.

Owing to the CMR effect, a magnetic field applied at $T_c$ would interact with the magnetic chains and by the way would enhance the conduction of orbital polarons along the chains with spin-charge-orbital and structural degrees of freedom coupled together. The small value of the interaction energy between the chains, $\Delta E \approx 3$ meV which stabilizes this local ground state agrees with predictions[1]. This nematic-like picture was actually predicted for cuprates[37]. In manganites, analytic models have indicated that the double exchange mechanism introduced by Zener[38] could not explain the effect of CMR. A charge-lattice coupling has been added to the double-exchange and it has been argued that the CMR effect was due to the self-trapping of the polarons at $T>T_c$[39]. Experimentally, this self-trapping does not occur[40]. A transition from a bipolaronic to a polaronic state has been also proposed[41] but it is not observed here. On the experimental side, we remind that the role of the orbital fluctuations of the type $3z^2-r^2$ has been previously shown in the spin dynamics of the narrow electronic band $Sm_{0.55}Sr_{0.45}MnO_3$ (Ref. 19). In this material the softening anomaly results from the 1$D$ paths of orbital polarons. It corresponds to the $x$=0.5 limit of the present study.

Actually most of experiments have been performed in compounds with a narrow electronic bandwidth such as $La_{1-x}Ca_xMnO_3$ in which the magnon anomalies are stronger than in $La_{1-x}Sr_xMnO_3$, and the magnetic excitations cannot be easily resolved into distinct energy levels. We emphasize that it is thanks to the observation of several energy levels and of their relation with phonon energies that the role of the orbital fluctuations in the magnetic excitations has been evidenced in present work. Within this analysis, not only the existence of ferromagnetic chains of orbital polarons has been revealed at $x_0$=0.2, but also the existence of an inter-polaron coupling with antiferromagnetic character at $x$=0.125. In spite of the observed differences between compounds with a large electronic band such as $La_{1-x}Sr_xMnO_3$ and those with a narrow band such as $La_{1-x}Ca_xMnO_3$, we argue that the underlying physics leading to the CMR property could be similar in both types of manganites at "equivalent" doping values.

In $La_{1-x}Ca_xMnO_3$ the optimum of CMR is expected to occur at $x$(Ca)=1/3. In the quasi-metallic state corresponding to the $x$(Ca)=0.17 and $x$(Ca)=0.2 doping values, a previous determination of the spin spectrum has revealed the existence of four energy levels in the [0.3-0.5] $q$-range[42]. We outline that these observations are similar to those reported at $x$(Sr)=0.125, which characterize 2$D$ orbital polarons. At larger $x$(Ca) values, the discreteness of the spectrum is smeared out so that the role of the phonons cannot be observed. In a first experiment with the nominal concentration $x$(Ca)=0.3, a broad magnetic excitation spectrum was reported at $E\approx$22.5 meV in the large $q$-range [0.3-0.5][18] resulting into a flattening when considering the whole $q$-dispersion. This behavior corresponds to the observations reported here at $x$(Sr)=0.3. Later, two other experiments have determined the spin dynamics at the same nominal $x$(Ca)=0.3 doping value but with a larger $T_c$ value[21,22]. The softening effect of the energies at the zone boundary was found to be nul[21] or very small[22] so that the whole $q$-dispersion can be described mainly by one phenomenological ferromagnetic $J_1$ coupling constant. These observations remind those reported here at $x_0$(Sr)=0.2. In the most recent study, an increase of the damping of the excitations was observed along Mn-O-Mn at E > 15 meV ($q$ > 0.25 rlu), interpreted as the effect of a phase separation picture[22]. Importantly, at the doping value $x$(Ca)=0.33, experiments by transmission electronic microscopy (TEM) have indicated the existence of small charge-ordered domains with however a too small size to be characterized[43]. In two other compounds, $La_{0.75}Sr_{0.25}MnO_3$ and $Pr_{0.67}Ca_{0.33}MnO_3$ small angle neutron scattering experiments with polarized neutrons have evidenced the existence of an inhomogeneous ferromagnetic state with a polymer structure or characterized by linear filaments[44,45]. This also reminds the description of ferromagnetic chains of orbital polarons provided here at $x_0(Sr)=0.2$. All the considerations given above suggest the existence of a similar charge-ordered state in different compounds at the doping with maximum magnetoresistance.

In summary, our study gives evidence for a spin dynamics characterized by a discrete magnetic energy spectrum $E^n_{mag}(q)$ and their direct relation with lattice dynamics in acoustic range. The $E^n_{mag}(q)$ spectrum is understood as arising from the internal excitations of "orbital polarons", which can be seen as a hole ($Mn^{4+}$) interacting with orbitals of the nearest adjacent $Mn^{3+}$ ions. The hopping mixes up dynamically all the possible orbital configurations of those surrounding $Mn^{3+}$ whose degeneracy is lifted by phonons. Furthermore, the shape of those polarons evolves with doping. At the transition between their $2D$ and their $3D$ evolution, interacting chains of polarons get formed and longitudinal acoustic excitations reveal the propagation of these polarons along the chains at $T \leq T_c$. This phase separation picture, between metallic and ferromagnetic chains in a non-metallic matrix may be crucial for CMR.

**Acknowledgements**
An important part of this experimental work has been done with F. Moussa and B. Hennion. They are gratefully acknowledged. The authors are also very indebted to S. Aubry, Y. Sidis, F. Onufrieva for many fruitful discussions. The work at the Peter the Great St. Petersburg Polytechnic University was supported by the Ministry of Education and Science of the Russian Federation (Grant No. 3.1150.2017/4.6).